# GRAVITY: metrology

Stefan Gillessen*[a], Magdalena Lippa[a], Frank Eisenhauer[a], Oliver Pfuhl[a], Marcus Haug[a], Stefan Kellner[a], Thomas Ott[a], Ekkehard Wieprecht[a], Eckhard Sturm[a], Frank Haußmann[a], Clemens F. Kister[a], David Moch[a], Markus Thiel[a]

[a]Max-Planck-Institute for extraterrestrial physics, 85748 Garching, Germany

**ABSTRACT**

GRAVITY is a second generation VLTI instrument, combining the light of four telescopes and two objects simultaneously. The main goal is to obtain astrometrically accurate information. Besides correctly measured stellar phases this requires the knowledge of the instrumental differential phase, which has to be measured optically during the astronomical observations. This is the purpose of a dedicated metrology system. The GRAVITY metrology covers the full optical path, from the beam combiners up to the reference points in the beam of the primary telescope mirror, minimizing the systematic uncertainties and providing a proper baseline in astrometric terms. Two laser beams with a fixed phase relation travel backward the whole optical chain, creating a fringe pattern in any plane close to a pupil. By temporal encoding the phase information can be extracted at any point by means of flux measurements with photo diodes. The reference points chosen sample the pupil at typical radii, eliminating potential systematics due differential focus. We present the final design and the performance estimate, which is in accordance with the overall requirements for GRAVITY.

## 1. INTRODUCTION

GRAVITY[1] is an adaptive optics-assisted, near-infrared fringe-tracking beam combiner instrument for a set of four telescopes from the VLTI[2]. During a typical observation, GRAVITY will provide simultaneous interferometry of two objects within the 2" field of view of the VLTI. If the two objects are stars, GRAVITY can work as a narrow-angle dual-beam interferometer for astrometry, but it is also possible to obtain phase-referenced, interferometric images for more complex sources. In either case, GRAVITY shall excel by its astrometric performance, allowing measurements of angles of order 10μas. The atmospheric phase residuals average out to less than 10μas error within few minutes, and hence the instrumental systematics should be under control at the same level. For the maximum baseline length of the VLTI of ≈ 100m, an angle of 10μas corresponds to an accuracy of the differential optical path length of 5nm, a fraction only of the observing wavelength 2μm. To this end, GRAVITY has a dedicated metrology system, the final design of which is presented here. The basic idea of a metrology system is to provide an optical link between the two interferograms of the two science beams, measured simultaneously. Interestingly, the end points of the metrology, i.e. the points to which the optical paths are measured, also define the astrometric baseline $B_A$ of the interferometer[3], that is the quantity that enters into the basic equation of dual beam interferometry (where the vector **s** is the difference between the two source vectors):

$$\Delta \mathrm{OPD} = \vec{B}_A \times \vec{s} + \Delta \mathrm{OPD}_{\mathrm{int}}$$

The specific requirements for the GRAVITY metrology are:

- Measuring quickly enough and, accurately: within 180s the OPD error should be ≈1nm.
- Stability: The measured OPD needs to be stable over at least 1 hour to ≈1nm.
- Cover the total length of the instrument and the VLTI up to points above M1, "in primary space".
- Sample the pupil with the metrology beams at representative locations.
- Provide a signal fast enough to stir the differential delay lines of the instrument.
- Follow all fast telescope and delay line motions, for all four telescopes in parallel.
- Require only minor retrofitting of the existing telescope and interferometer hardware.

*ste@mpe.mpg.de; phone 49 89 30000 3839

## 2. WORKING PRINCIPLE

The basic working principle is the same as described for the preliminary design[4,5]. A laser beam is injected backwards at the two beam combiners, in a way that one of the two paths includes a phase shifter by which the relative delay between the two beams can be adjusted. For each telescope one gets hence such a pair of beams, which are point sources in the field and thus completely overlap in pupil planes. Since they originate from the same laser source, the two beams are coherent and thus the overlay yields an interferogram, the phase of which encodes the difference in optical path (figure 1). A convenient way to measure the phase is phase shifting interferometry, which can be achieved here by means of the phase shifter. The actual detection of the interferogram happens after the light has travelled backward through the instrument, the full VLTI train, and the telescope, after having been reflected by the primary mirror of the telescope: The spider arms of the telescopes (carrying the secondary mirror) will be equipped with photodiodes looking downward onto M1. Such a diode measures the intensity of the metrology interferogram at its corresponding location in the pupil, and by phase shifting interferometry a phase can be obtained from that signal.

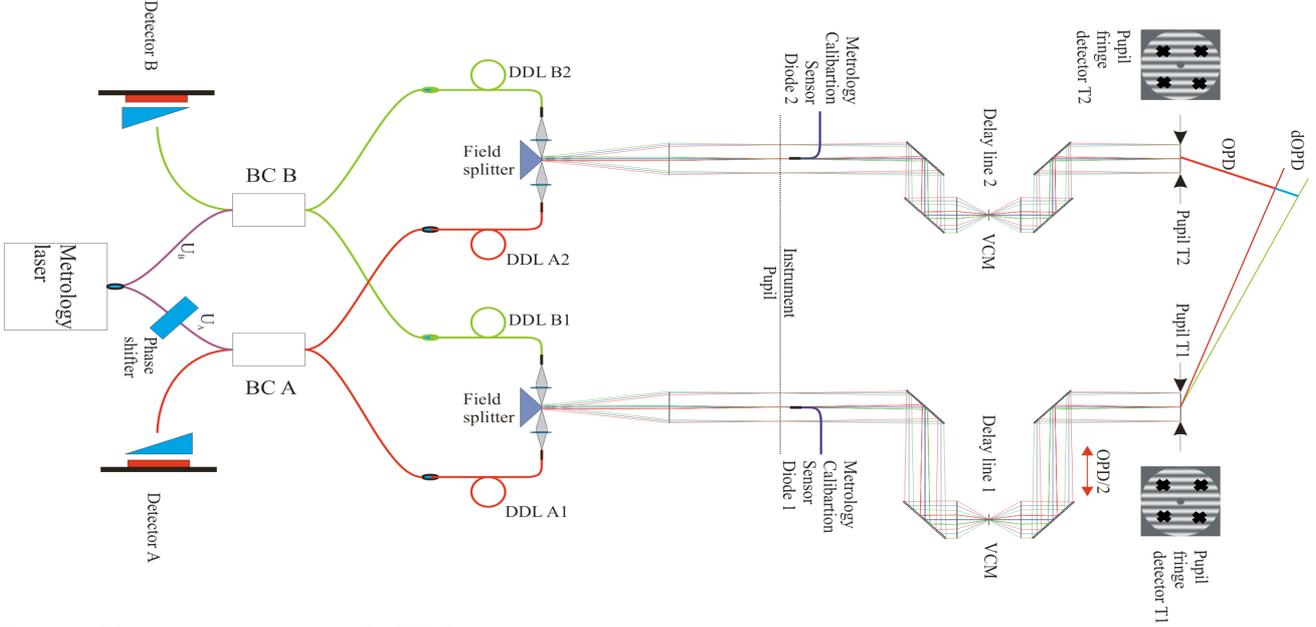

Figure 1: Working principle of the GRAVITY metrology.

This system thus measures the optical path difference between the two beams to each telescope. In the above equation of dual beam interferometry, however, the optical path difference between the two telescopes for each beam occurs. Nevertheless, one can convince oneself, that the relevant information is measured: For both beams A and B the phase between telescopes k and l at their locations T are given by

$$\phi_{kl,A} = (\vec{T}_k - \vec{T}_l).\vec{s}_A + i_k - i_l$$
$$\phi_{kl,B} = (\vec{T}_k - \vec{T}_l).\vec{s}_B + j_k - j_l$$

where i and j denote the internal delays. The metrology measures

$$m_k = i_k - j_k$$

which occurs in the phase formula after re-grouping the terms:

$$\Delta\Phi_{kl} = (\vec{T}_k - \vec{T}_l).(\vec{s}_A - \vec{s}_B) + m_k - m_l$$

The basic requirement on the phase accuracy is given by the astrometry, i.e. the 1nm error should be reached within 180s. This corresponds to 13nm/√s or for a frequency of 1kHz to roughly 400nm per phase shifting cycle. Another requirement is that a phase jump larger than π must never occur. For the expected number of operating hours for GRAVITY and assuming Gaussian errors, this yields an accuracy requirement of roughly 200nm per cycle. Note that also the frequency of the metrology cycle has a (soft) requirement, originating from acquisition procedure: During the latter, the fibers in GRAVITY have to be moved by up to ±6mm in optical path, which has to be followed by the metrology - at a rate of 1kHz this yields a preset time of roughly 25s, which is still acceptable.

## 3. IMPLEMENTATION

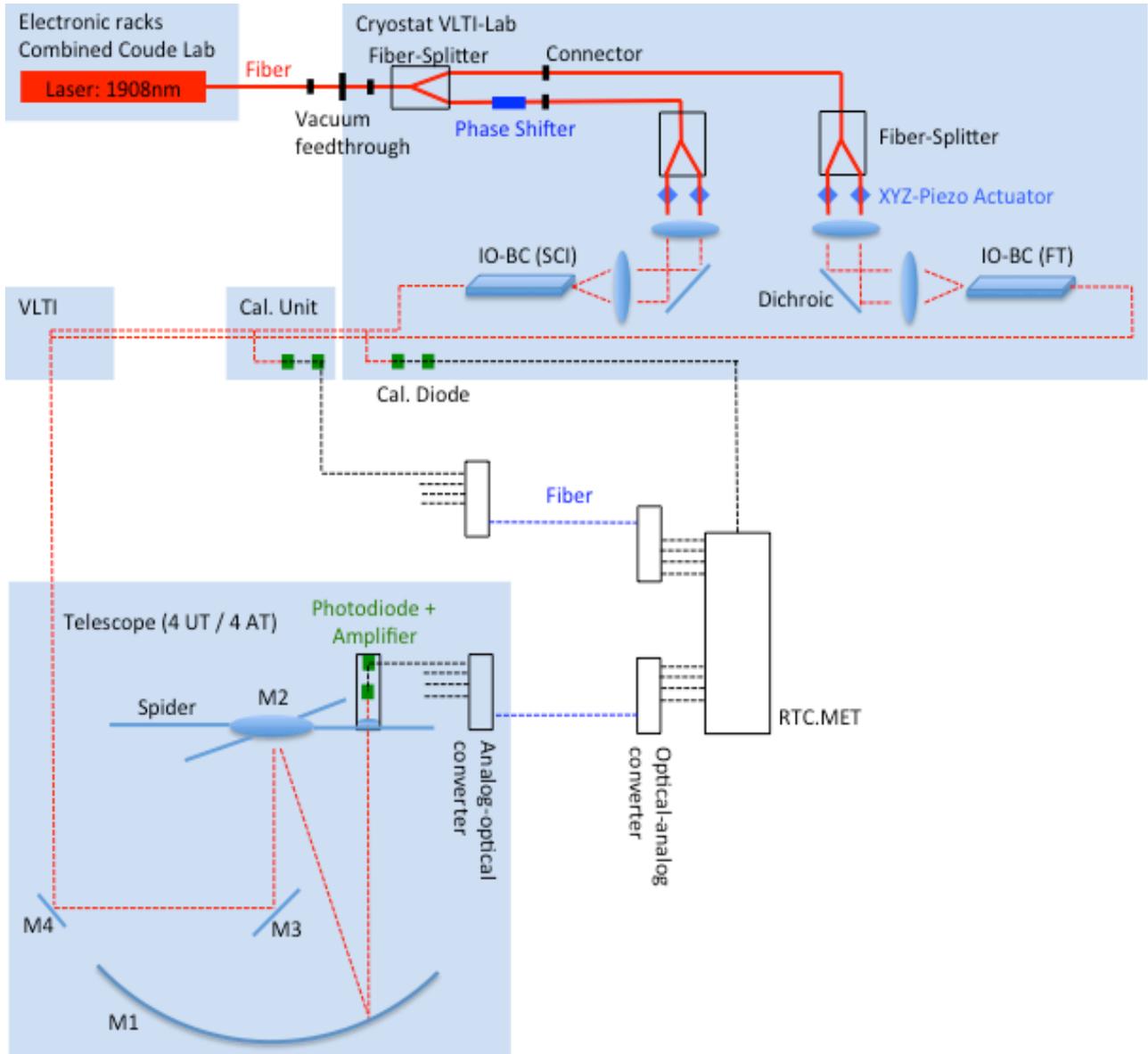

Figure 2: Schematic view of the implementation of the GRAVITY metrology

The implementation of the GRAVITY metrology is illustrated in figure 2. A fibred laser source with $\lambda=1908$nm is feeding a fiber optics entrance splitter. After the split, in one arm an electro-optical phase shifter is included. Both arms then feed a respective fiber optics exit splitter. These are equipped with exit fibres mounted on XYZ translation stages. By means of bulk optics the fiber exits are projected onto the integrated optics beam combiner[6]. Note that the injection cannot be done in fibers only, since one would block then the science light of the corresponding IO output (figure 4). The overlay of the science and metrology beams is achieved by means of a dichroic mirror mounted at an angle of 45°. From the beam combiner on, the path of the metrology light is identical to the one of the science light up to the primary mirror of the telescopes. The only path, which is non-common between science and metrology light, is the part from the exit splitters to the beam combiner, i.e. a few cm in the exit fibers and the bulk optics part. The design must guarantee that this non-common path is kept stable between each pair of exit fibers to the nm-level over the course of an observation. Special care also has to be taken to block back-reflections of the metrology light onto the detectors[7].

At the telescopes, each of the fiber arms is equipped with a photodiode looking into the metrology light reflected at the primary mirror. The devices consist of a collector lens, a 1908nm filter, a sensitive photodiode and a custom-developed amplifier circuit converting the expected level of few nW of incident light to a voltage between 0 and 10V. This voltage is then transported to the VLTI laboratory (where the work stations of the instrument are located) by means of an analog-optical fibre link.

Similar diodes are mounted also inside the instrument at the fiber coupling units[8], allowing for simultaneous monitoring of the GRAVITY-internal delays (without the VLTI beam train contributions). This quantity is relevant for stirring the differential delay lines of GRAVITY. Finally, also the calibration unit of GRAVITY carries such photodiodes, with the aim of simulating the telescopes in every respect.

## 4. COMPONENTS

### 4.1 Laser source

The GRAVITY metrology laser is a 2W continuous wave laser operating at 1908nm stabilized to ± 30MHz and a line width < 10MHz. It was built by Menlo Systems as a turn-key solution. The laser is in routine operation at MPE. Its wavelength is ultimately the ruler by which GRAVITY realizes its astrometric performance.

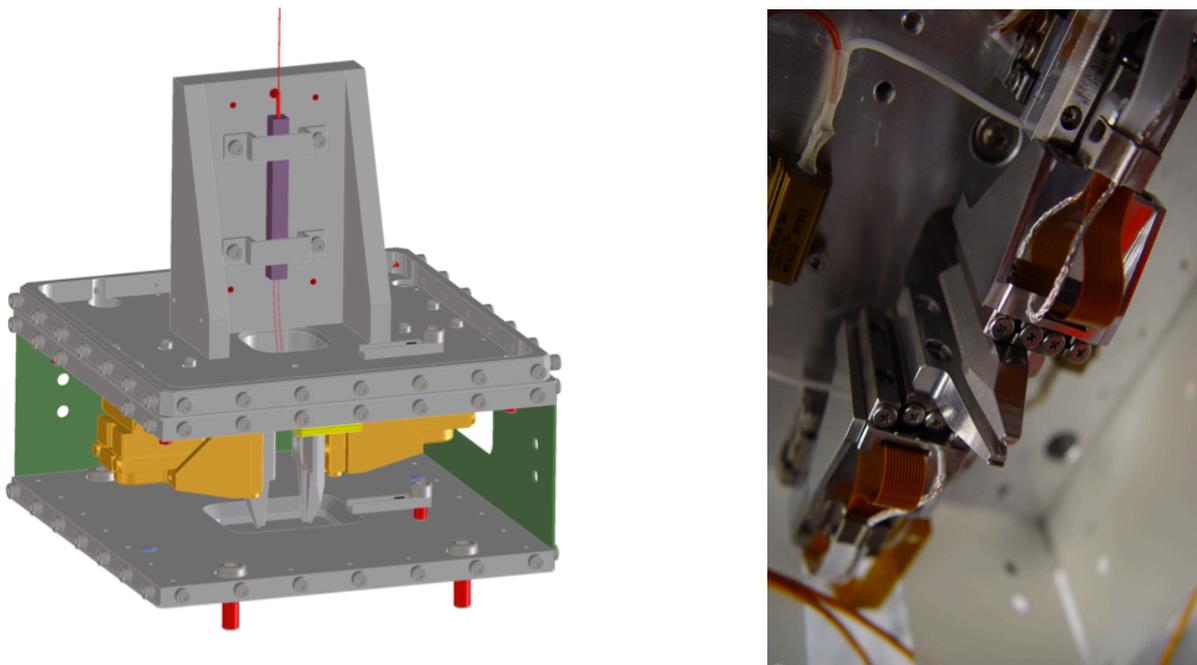

Figure 3: Left: Schematic view of the fiber injection unit with the fiber feeding the splitter from the top. Right: Photo of the fiber positioner prototype, showing the mounts into which the fibers are glued.

## 4.2 Splitters

The splitters are custom-tailored integrated optics splitters with a nominal operating wavelength at 1550nm, which we have validated for operation at 1908nm. The entrance splitter is critical for its power level of up to 2W. A device from AMS fiber technology has proven to be suitable. Due to the ≈50% transmission of the phase shifter in one of the arms, the entrance splitter actually needs to split with a power ratio of 2:1, such that both exit splitters receive the same power level. The latter devices are procured by LEONI fiber optics. They are equipped on the exit side with two bare fibers of equal length (to within 1mm), such as to minimize optical path difference variations due to thermal variations. (For the given length tolerance and a total length of a few cm stabilizing thermally to mK brings the optical path difference in accordance with the error budget.)

## 4.3 Phase shifter

The phase shifter is a commercial device from Photline, allowing to electro-optically shift the phase of the light passing by up to 2λ when applying voltages between -10V and +10V. It can operate in the MHz regime, such that it suits the kHz operation required here. The device is currently being tested at MPE.

## 4.4 XYZ translation stages

The fiber ends are glued onto piezo translation stages from SmarAct. Three linear devices together form one stage, allowing moving the fiber ends in 3D by a few mm. These degrees of freedom are used to couple the light from the fibers to the (science) outputs of the integrated optics beam combiner. This component has been prototyped and tested at MPE (figure 3 & 4).

## 4.5 Injection optics

The injection optics are part of the GRAVITY spectrometers[9]. It contains a first set of lenses to create a collimated beam, which then passes narrow band 1908nm filters to block the unwanted broad laser emission. After that, the beam hits a dichroic mirror, reflecting the metrology light by 90° onto two of the integrated optics outputs. The science light (2.0μm - 2.4μm) leaves the exact same outputs and is transmitted by the dichroic (figure 4).

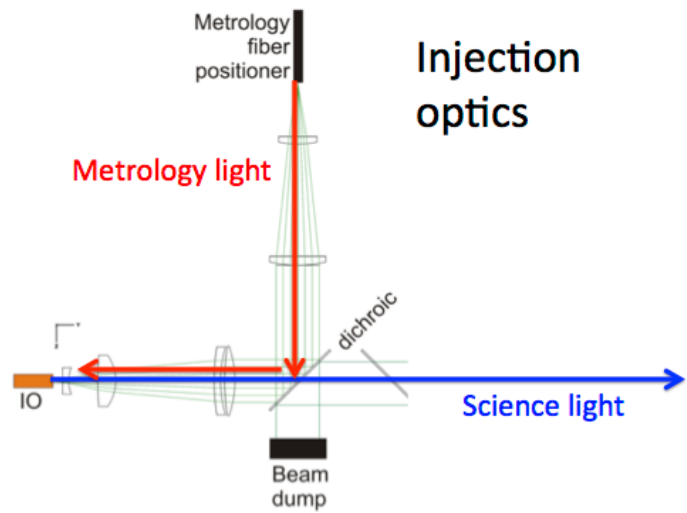

Figure 4: Optical layout of the injection unit.

## 4.6 Receivers

The receivers are photodiodes from Hamamatsu with a sensitivity of 1A/W and low noise. Given the expected signal level, a collector lens of a few cm diameter is used to gather sufficient light and to focus the light onto the diode. On the outside, a 1908nm line filter cleans the incident light spectrally. The diode signal is amplified by a MPE-built two-stage amplifier circuit. The whole unit is mounted in a compact aluminum box, the electrical shielding of which is crucial for the low-noise performance.

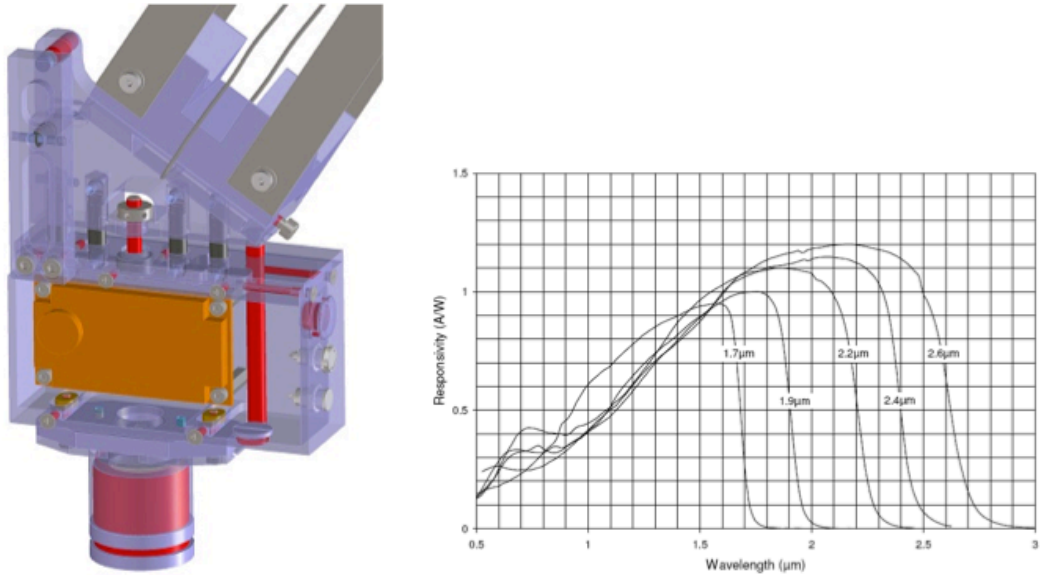

Figure 5: Left: Mechanical design of the receiver housing. The downward looking tube contains the collecting lens and a filter, the metal box houses the amplifier circuit. Right: Sensitivity curves for IR photodiodes. A cut-off of 2.2μm yields optimum sensitivity at the metrology wavelength of 1.9μm.

### 4.7 Signal transmission

The four voltages per telescope need to be fed back to the control computers of the instrument. Given the distances of a few hundred meters, it is most practical to convert the analog signals to digital ones, which then can be transported with standard glass fibres. At the backend, the signals are converted back to the original analogue voltage and read in by standard hardware. Such transmission systems are commercially available (the frequency of the metrology signal is comparable to audio signals), and we purchased one from MK Messtechnik.

## 5. SYSTEMATICS

There are two main sources of systematic errors, both of which we addressed experimentally.

### 5.1 Non-common path in the injection unit

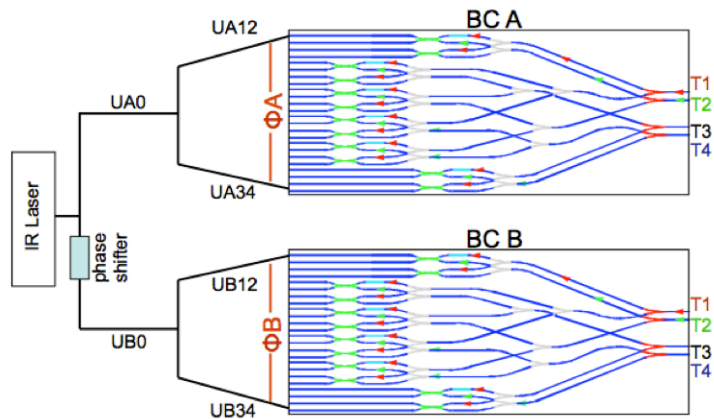

Figure 6: The non-common path, which enters the astrometric error budget, is $(\Phi_A - \Phi_B)$ for the chosen injection scheme. Since the phase differences $\Phi_A$ and $\Phi_B$ occur in the two distinct spectrometers, each of them has to be kept stable to the nm-level.

For the chosen injection scheme, a few cm of non-common path between metrology and science light exist, that impact the astrometric error budget (figure 6). The measured phases are:

$\Phi_1 = UA0 + UA12 - UB0 - UB12 + ...$

$\Phi_2 = UA0 + UA12 - UB0 - UB12 + ...$

$\Phi_3 = UA0 + UA34 - UB0 - UB34 + ...$

$\Phi_4 = UA0 + UA34 - UB0 - UB34 + ...$

The corresponding astrometric relevant quantities are:

$\text{dOPD}12 = \Phi_1 - \Phi_2 = \qquad\qquad\qquad\qquad\qquad\qquad\qquad\qquad 0 + ...$

$\text{dOPD}13 = \Phi_1 - \Phi_3 = \quad (UA12 - UA34) - (UB12 - UB34) + ... = \Phi_A - \Phi_B + ...$

$\text{dOPD}14 = \Phi_1 - \Phi_4 = \quad (UA12 - UA34) - (UB12 - UB34) + ... = \Phi_A - \Phi_B + ...$

$\text{dOPD}23 = \Phi_2 - \Phi_3 = \quad (UA12 - UA34) - (UB12 - UB34) + ... = \Phi_A - \Phi_B + ...$

$\text{dOPD}24 = \Phi_2 - \Phi_4 = \quad (UA12 - UA34) - (UB12 - UB34) + ... = \Phi_A - \Phi_B + ...$

$\text{dOPD}34 = \Phi_3 - \Phi_4 = \qquad\qquad\qquad\qquad\qquad\qquad\qquad\qquad 0 + ...$

Hence, for four of the six baselines, the difference of the phases $\Phi_A$ and $\Phi_B$ is relevant. It is crucial that this path difference is kept stable to the nm-level during the observations. Given that $\Phi_A$ and $\Phi_B$ occur at two different spectrometers and are thus uncorrelated, one actually needs to stabilize both ΦA and ΦB individually. In practice, this means that the light paths from the exit splitters to the beam combiner need to be stable, which includes a fiber part and a bulk optics part. The fiber part yields an OPD difference of

$$\text{OPD}_T = 8.71157 \times 10^{-6}\,\text{K}^{-1}(L_0 \delta T + \delta L \Delta T + \delta L \delta T)\quad,$$

where $L_0$ is the length of the fiber (≈5cm), $\delta T$ is the temperature difference between the two fibers (≈mK), $\delta L$ is the difference in the fiber lengths (≈1mm) and $\Delta T$ is the change in temperature since the zero-point calibration of the metrology (≈ 10mK). The numerical constant follows from the material constants of fused silica, where both the thermal expansion coefficient and the change of refractive index with temperature need to be taken into account. With these numbers, the change in OPD remains within the error budget.

The bulk optics part is harder to quantify, and we have started testing the design by means of prototyping. In particular, we are testing whether polarization state changes can affect the phase and whether the free beam injection is stable in all three dimensions.

### 5.2 Differential aberration in the beam train

The astronomical interferograms yield a phase, which is the average over the whole pupil, weighted with the Gaussian field distribution of the fiber mode projected into the pupil plane. Since the receivers sample the phase only at a finite number of points in the pupil, changes in the differential aberrations of the two beams might introduce phase errors. Hence, the question arises, how to optimally place the receiver diodes. Intuitively it is clear that mounting them point-symmetrically to the pupil center will remove any tip/tilt wavefront error contributions. Also the next Zernike order (focus) can be removed, since one has the freedom to choose the radius at which the diodes are mounted. The differential focus term actually also is expected to be a dominant contribution, since it is only badly controlled for the fibers used in GRAVITY for spatial filtering, OPD control and polarization adjustment. One searches the radius $r_0$ where the weighted pupil average of the Zernike term equals the value of the Zernike term itself, i.e.

$$Z_4(r_0, \theta) = \int_0^1 Z_4(r, \theta)\, w(r)\, r\, dr \bigg/ \int_0^1 w(r)\, r\, dr\quad.$$

This yields $r_0 = 0.645$ (where the pupil radius is normalized to 1 and w is the weight function).

Any higher orders cannot be corrected additionally, except by increasing the number of diodes beyond 4. Hence, it needs to be checked that the higher order terms are small enough to be neglected. To this end we have conducted a test campaign at the VLTI, bringing a test interferometer to the VLTI laboratory and shining with a beam simulating the metrology light all the way to the telescope. During this test campaign, we have mounted a camera on the M1 mirror cell, by which we were able to observe the resulting interferogram on M2 (i.e. the full pupil) in scattered light. Such an interferogram contains the desired information, i.e. it can be unfolded to show the phase and its deviation from a pure tilt (figure 7). The result was that indeed we were not able to see any higher order terms beyond a focus component.

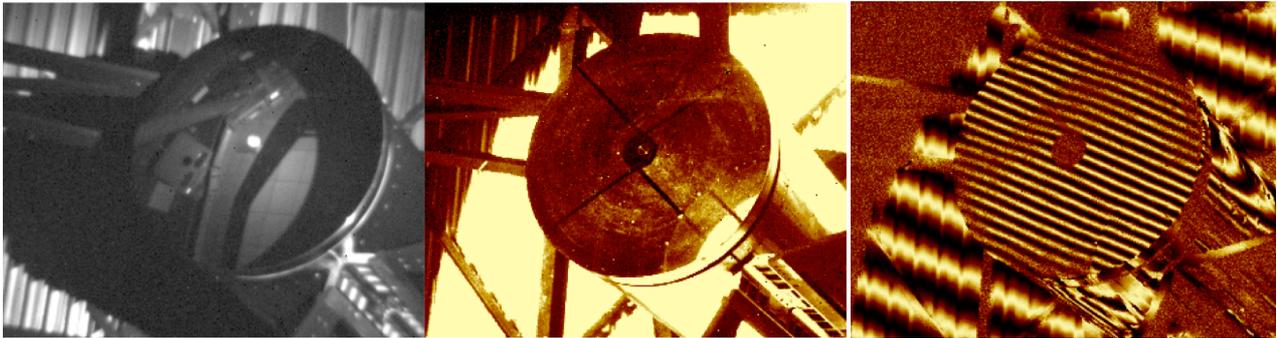

Figure 7: Testing the differential aberrations of the VLTI beam train. Left: view of the secondary mirror from below. Middle: The mirror is illuminated with metrology light from the VLTI lab, showing the spider arms and the central obscuration from a mask on M8. Right: Interferogram on the secondary mirror from which the aberrations can be determined.

## 6. PERFORMANCE

### 6.1 Power budget

For a nominal input power of 2W, the resulting power reaching the primary mirror is below 1mW - a result of the numerous optical elements with transmission numbers below unity, the splitting of the light to four telescopes and the overall VLTI + GRAVITY transmission. Only a small fraction of the light (the 1mW is distributed in a Gaussian fashion over the full pupil) is captured by the collector lenses, yielding a power level of ≈1nW per diode. Hence, the amplifiers need to convert the resulting photocurrent of 1nA to a measurable voltage. The noise of the amplifier (dominated by the shunt resistance and shot noise) can be estimated to $\approx 5 \times 10^{-11}$ W (a number which were also able to verify in a laboratory test), or a SNR > 10 for an individual read of the voltages. For the ABCD algorithm,

$$d\phi^2 = \frac{dI^2}{2I_0^2}$$

This yields dΦ < 0.07 or dOPD < 135nm, i.e. consistent with the requirement of 200nm accuracy per single read.

### 6.2 OPD budget

The following table gives the metrology astrometry error budget:

| Phase error contribution | requirement / design value | Phase error, 3 min, 2", UT [nm] | Phase error, 3 min, 6", AT [nm] |
|---|---|---|---|
| **Wavelength uncertainty IR laser** | 30Mhz | 0.220 | 1.110 |
| **Intensity noise IR laser** | 0.5% in 30ms | 0.014 | 0.014 |
| **Phase shifter amplitude modulation** | 0.30% | 0.640 | 0.640 |
| **Phase shifter calibration + stability** | λ/5000 | 0.380 | 0.380 |
| **Amplitude modulation IO coupling jitter** | 0.02% | 0.040 | 0.040 |
| **Amplitude modulation in VLTI train** | 1% / sec. | 0.160 | 0.160 |
| **Metrology zero-point calibration + stability** | 0.5nm | 0.500 | 0.500 |
| **Metrology receiver noise** | 30nm / ms | 0.075 | 0.038 |
| **Quasi-static aberrations** | measured | 0.450 | 0.450 |
| **Non-common path** | estimate | 0.530 | 0.530 |
| **Total** | | 1.272 | 1.673 |

The main contributors are the phase shifter calibration and repeatability, the non-common path errors and the quasi-static aberrations. Overall, the error budget fulfills the requirements for GRAVITY - the phase error introduced by the metrology is well below 5nm.

## 7. CONCLUSIONS

The metrology system of GRAVITY presented here minimizes the systematic astrometric errors related to the internal optical paths. It fulfills the requirements for 10μas-level astrometry. The system should be able to measure the optical paths with nm-level precision. Also, it can run fast enough, providing feedback for the GRAVITY-internal fibered differential delay lines.